\documentclass[journal]{IEEEtran}

\ifCLASSINFOpdf
\else
   \usepackage[dvips]{graphicx}
\fi

\usepackage{graphicx}
\usepackage{amsmath,amsfonts}
\usepackage{textcomp}
\usepackage{stfloats}
\usepackage{url}

\usepackage{wasysym}
\usepackage{multirow}
\usepackage{arydshln} 
\usepackage{subcaption}
\usepackage{xcolor}
\usepackage{cite}
\usepackage[colorlinks=true, urlcolor=blue]{hyperref}
\def\BibTeX{{\rm B\kern-.05em{\sc i\kern-.025em b}\kern-.08em
    T\kern-.1667em\lower.7ex\hbox{E}\kern-.125emX}}
\usepackage{balance}

\usepackage{xcolor}
\newcommand{\seqit}[1]{\bar{#1}}

\newcommand{\x}{\ensuremath{\mathbf{x}}}

 \newfont{\msym}{msbm10}

\renewcommand{\sc}{\seqit{c}}

\newcommand{\z}{\ensuremath{\mathbf{z}}}

\newcommand{\conv}{HyperSpotter}
\newcommand{\wconv}{HyperSpotter-w~}
\newcommand{\cconv}{HyperSpotter-c~}

\newcommand{\kc}{\mathbf{k}}  
\newcommand{\Wkc}{\mathbf{W}^{\kc}}  

\begin{document}

\title{Keyword Spotting with Hyper-Matched Filters for Small Footprint Devices}


\author{
Yael Segal-Feldman\IEEEauthorrefmark{1},
Ann R. Bradlow\IEEEauthorrefmark{2},
Matthew Goldrick\IEEEauthorrefmark{2}, and
Joseph Keshet\IEEEauthorrefmark{1}
\\
\IEEEauthorrefmark{1}Technion – Israel Institute of Technology, Faculty of Electrical and Computer Engineering, Israel \\
\IEEEauthorrefmark{2}Northwestern University, Department of Linguistics, USA
\thanks{The corresponding author is Yael Segal-Feldman (email: segal.yael@campus.technion.ac.il). This document is a pre-print version of the paper.}
}
\maketitle

\begin{abstract}
Open-vocabulary keyword spotting (KWS) refers to the task of detecting words or terms within speech recordings, regardless of whether they were included in the training data. This paper introduces an open-vocabulary keyword spotting model with state-of-the-art detection accuracy for small-footprint devices. 
The model is composed of a speech encoder, a target keyword encoder, and a detection network. The speech encoder is either a tiny Whisper or a tiny Conformer. The target keyword encoder is implemented as a hyper-network that takes the desired keyword as a character string and generates a unique set of weights for a convolutional layer, which can be considered as a keyword-specific matched filter. The detection network uses the matched-filter weights to perform a keyword-specific convolution, which guides the cross-attention mechanism of a Perceiver module in determining whether the target term appears in the recording. The results indicate that our system achieves state-of-the-art detection performance and generalizes effectively to out-of-domain conditions, including second-language (L2) speech. Notably, our smallest model, with just 4.2 million parameters, matches or outperforms models that are several times larger, demonstrating both efficiency and robustness.
\end{abstract}

\begin{IEEEkeywords}
Keyword spotting, Spoken term detection, Open vocabulary, Small footprint device, Hypernetwork
\end{IEEEkeywords}

\IEEEpeerreviewmaketitle
\section{Introduction}
\label{sec:intro}

Keyword Spotting (KWS) is the task of identifying a predefined set of target keywords. It is an essential task in speech recognition systems, particularly crucial for voice assistants on \emph{small-footprint devices}, such as smartphones and small speakers \cite{lopez2021deep}. Throughout the years, various techniques have been explored for this task, beginning with basic deep neural networks based on Convolutional Neural Networks (CNNs) \cite{chen2014small, sainath2015convolutional} to Recurrent Neural Networks (RNNs) as audio encoders \cite{shan2018attention,arik2017convolutional}.

As research has advanced, attention has increasingly shifted toward \emph{open vocabulary} KWS, which focuses on detecting keywords, terms, or phrases in speech recordings --- even when these targets were \emph{not} seen during training\footnote{Throughout this paper, we use the term \emph{target keyword} to refer to a word, term, or phrase of interest.}. A central challenge in this task is achieving the flexibility required to accurately detect previously unseen keywords within spoken utterances. Various approaches have been proposed to address this challenge. One widely adopted strategy is query-by-example (QbE), which leverages enrolled audio samples of the target keyword to facilitate detection \cite{chen2015query, lugosch2018donut, huang2021query, rikhye21_interspeech, r22_interspeech}. In a QbE approach, the system builds a reference model using enrolled audio examples of the target keyword. This model is then used to match incoming audio, identifying instances of the target keyword based on similarity to the stored examples. An alternative QbE method leverages Text-to-Speech (TTS) technology to synthesize audio for new keywords \cite{vuppala2024open}, which serves as the reference for detecting those keywords in future recordings. However, the performance of QbE approaches is highly sensitive to variations in recording conditions, making consistency in the audio environment crucial for reliable detection. To overcome this problem, several studies have focused on deep acoustic models trained with Connectionist Temporal Classification (CTC) loss \cite{graves2006connectionist} and techniques for detecting the target keywords within speech \cite{zhuang2016unrestricted, xuan2019robust,bluche2020small,wei2021end}. 

Other research approaches involve presenting the target keyword in phonetic or graphemic form and using the resulting keyword embedding as input to the acoustic model \cite{keshet2009discriminative, sacchi2019open, fuchs2021cnn, shin2022learning, nishu2023flexible, nishu2023matching, pudo2024improved}. This method translates the target keyword into a representation that the acoustic model can process, allowing the system to recognize the keyword based on its phonetic or graphemic characteristics. Additionally, some methods utilize hyper-networks, where a primary model receives the target keyword in either phonetic or graphemic form and subsequently generates weights for another network. This approach effectively creates a tailored set of parameters for the secondary network based on the input keyword. For example, \cite{bluche2020predicting} employs this technique to produce weights specifically for a single CNN layer, enhancing its ability to detect the target keyword. On the other hand, \cite{navon2023open} generates weights for the normalization layer of a Transformer \cite{vaswani2017attention}, adjusting the model's internal parameters to better accommodate the target keyword. 

Despite these efforts, the mentioned works still face a major challenge: their performance dramatically degrades when converted to work on small-footprint devices. The models of \cite{bluche2020predicting, shin2022learning, nishu2023flexible, nishu2023matching} are designed for small-footprint devices but exhibit reduced performance on challenging target keywords. On the other hand, \cite{navon2023open} demonstrates excellent performance, but their model is too big for small-footprint devices.

This paper presents an open-vocabulary KWS model suitable for small-footprint devices with a state-of-the-art detection rate. Inspired by the works of \cite{navon2023open} and \cite{bluche2020predicting}, the proposed system is composed of a speech encoder, target keyword encoder, and detection network. The speech encoder provides an effective speech representation. 
As part of our effort to address the small-footprint challenge, the encoder is implemented using either tiny Whisper \cite{radford2023robust} or as tiny Conformer \cite{nishu2023flexible}. 

The target keyword encoder learns to represent an unseen keyword. It functions as a hyper-network, and rather than representing the keyword as an embedding vector, it generates a unique set of weights for a convolution layer of the detection network. The detection network leverages these weights to perform keyword-specific detection. It is implemented with a Perceiver architecture \cite{jaegle2021perceiver} that allows the use of Transformer layers with fewer parameters, making it suitable for small-footprint devices. Specifically, the keyword-specific encoder's weights are used in the convolutional operation to guide the cross-attention weights within a Perceiver module. The overall detection network can be considered as a \emph{matched filter} that highlights keyword locations. As a result, the audio encoder output is directed towards the target keyword-specific, enabling the detection of whether the keyword-specific was uttered in the recording. Our implementation and trained models are available on GitHub\footnote{\url{https://github.com/YaelSegal/HyperSpotter}}.

We performed an empirical evaluation of the model across various model sizes and multiple datasets. The evaluation was performed for in-domain scenarios, where both the training and testing data originate from the same source. Furthermore, we assessed the model's performance in out-of-domain generalization scenarios, including out-of-domain datasets, languages unseen during training, and non-native English speakers. The results indicate that our system achieves a state-of-the-art detection rate and generalizes effectively to out-of-domain scenarios. Additionally, our smallest network, with 4.2M parameters, performs comparably to a model with 10M parameters in certain scenarios.

The contributions of our work are as follows: (1) We propose a novel, state-of-the-art open-vocabulary KWS model with an architecture suitable for small-footprint devices; (2) We evaluate several implementations of the proposed architecture using various audio encoders; (3) We present results across multiple datasets and diverse out-of-domain scenarios, including evaluation on L2 speakers; and (4) Our system's code is openly accessible for academic use, setting it apart from many similar KWS projects.

This paper is structured as follows. Section \ref{sec:model} introduces our proposed method and model architecture. Section \ref{sec:experiments} details the datasets used and the experimental setup and presents results on the datasets. Finally, Section \ref{sec:conclusion} concludes with remarks and discusses future directions.

\section{Related Work}
\label{sec:realted_work}


We review related work and organize it according to its relevance to the core ideas presented in this paper.

\subsection{Keyword Spotting for Small-Footprint Devices}

Keyword Spotting is a crucial task, especially for small-footprint devices like smartphones and small speakers, which power voice assistants such as Amazon's Alexa, Apple's Siri, Microsoft's Cortana, and Google's ``Hey Google'' \cite{sainath2015convolutional, lopez2021deep, nishu2023matching}. Early attempts to incorporate deep neural networks into keyword spotting employed basic CNN architectures \cite{chen2014small, sainath2015convolutional} and TDNN combined with Hidden Markov Models (HMM) \cite{sun2017compressed}. These efforts progressed to more advanced designs, including CNNs with residual layers \cite{tang2018deep}, depthwise convolutions \cite{xu2020depthwise}, multi-scale temporal convolutions \cite{li2020small}, sinc-convolutions \cite{mittermaier2020small}, and slimming CNNs \cite{akhtar2023small}. Beyond CNN-based approaches, later methods incorporated RNNs as audio encoders \cite{arik2017convolutional} and integrated attention-based mechanisms \cite{shan2018attention}.
Another line of research explores contextual biasing of ASR models toward known keywords \cite{michaely2017keyword} and hot-fix keyword recognition using model reprogramming techniques \cite{ku2024hot}. Despite these advancements, all these methods are constrained to spotting pre-defined keywords.

\subsection{Open Vocabulary Keyword Spotting}

Open Vocabulary Keyword Spotting involves identifying target keywords within audio recordings, even if those keywords were not included in the training data. A significant advantage of these systems is their ability to allow users to easily add new keywords or commands without retraining the entire model. This feature is particularly valuable in applications where new terms or commands need to be integrated quickly and efficiently, facilitating the development of more dynamic and adaptable voice-controlled systems \cite{zhuang2016unrestricted, trmal2017kaldi, lugosch2018donut}.

In the early efforts to address open vocabulary KWS based on full ASR outputs, \cite{zhuang2016unrestricted} proposed a method that involved training an audio encoder using an LSTM architecture with CTC loss. They employed a minimum edit distance algorithm to match the phonetic sequence of the keyword within the LSTM output, aiming to identify the presence of the keyword. Expanding on this approach, \cite{bluche2020small} introduced a CTC-based KWS model designed for small-footprint devices. They achieved this by quantizing the LSTM audio encoder to reduce its size while preserving its performance, making it more suitable for deployment on resource-constrained devices.

Building on this progress, \cite{he2017streaming} introduced a keyword detection system that uses an RNN-T model as the audio encoder. Their method features an attention mechanism designed to focus the model on the keyword, with the model generating ``start-of-keyword" and ``end-of-keyword" tokens to indicate the presence of the keyword in the audio. Similarly, \cite{wei2021end} also utilizes ``start-of-keyword" and ``end-of-keyword" tokens but employs CTC loss. Their approach integrates a cross-attention mechanism to guide the audio encoder toward the keyword and accurately locates the keyword within the audio.

Performing full transcription for keyword spotting can be overly complex, leading to alternative approaches that focus on keyword spotting using keyword characters or phoneme embeddings. \cite{sacchi2019open, shin2022learning, nishu2023flexible, nishu2023matching} proposed encoding both the audio and the keyword into a shared embedding space using various audio and keyword encoders. They then developed methods to determine keyword presence, ranging from distance metrics to additional deep neural network components. In contrast, \cite{fuchs2021cnn} and \cite{lee23d_interspeech} introduced a model that encodes audio while receiving keyword embeddings and then assesses whether the keyword was spoken in the audio.

\subsection{Hypernetworks}
The most pertinent work related to our approach involves the application of hypernetworks for KWS. Hypernetworks \cite{klein2015dynamic, ha2016hypernetworks} are advanced neural networks designed to generate weights for a secondary target network. The core innovation of hypernetworks is their ability to adapt the output weights based on the input provided to them, enabling dynamic adjustment of the target network's parameters. This technique has demonstrated substantial potential across various domains, including meta-learning and few-shot learning \cite{sendera2023hypershot, zhao2020meta}, language modeling \cite{suarez2017language, ruiz2024hyperdreambooth}, and multi-objective optimization \cite{navon2020learning}. In the context of KWS, hypernetworks learn parameters that represent the keyword, which are then used by the target network to detect the keyword. Bluche et al. \cite{bluche2020predicting} introduced an approach where an LSTM-based hypernetwork outputs weights for the final CNN layer of an audio encoder, with the hypernetwork receiving the keyword as input. Conversely, \cite{navon2023open} applied a hypernetwork to generate weights for the normalization layer of a Transformer, adjusting the model’s internal parameters to emphasize the keyword.

\begin{figure*}[h]
    \centering
    \begin{subfigure}{0.5\textwidth}
        \centering
        \includegraphics[width=1\linewidth]{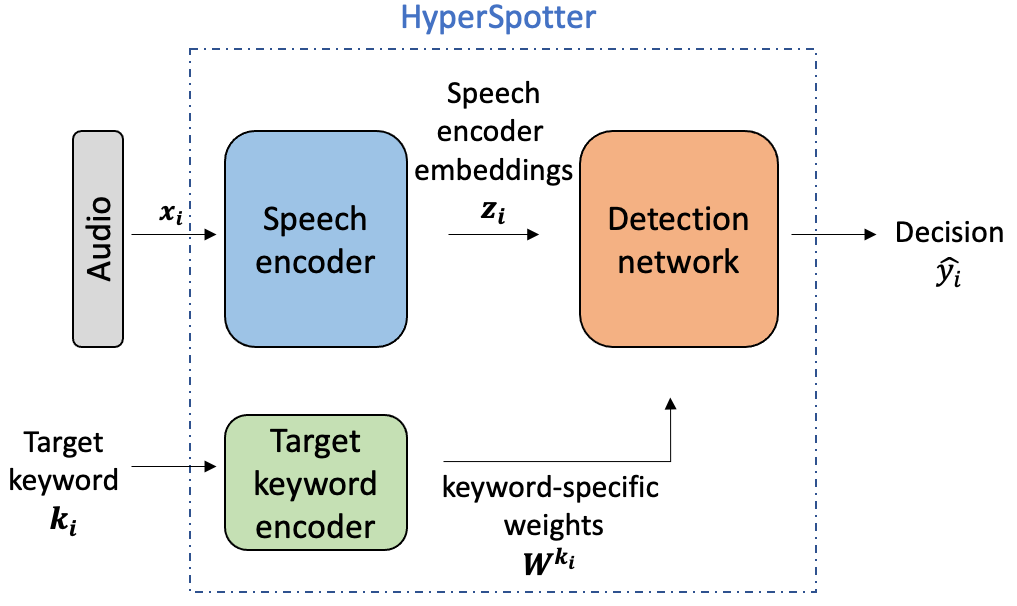}
        \caption{}
        \label{fig:sub1}
    \vspace{0.5cm}
    \end{subfigure}
    \begin{subfigure}{0.8\textwidth}
        \centering
        \includegraphics[width=\linewidth]{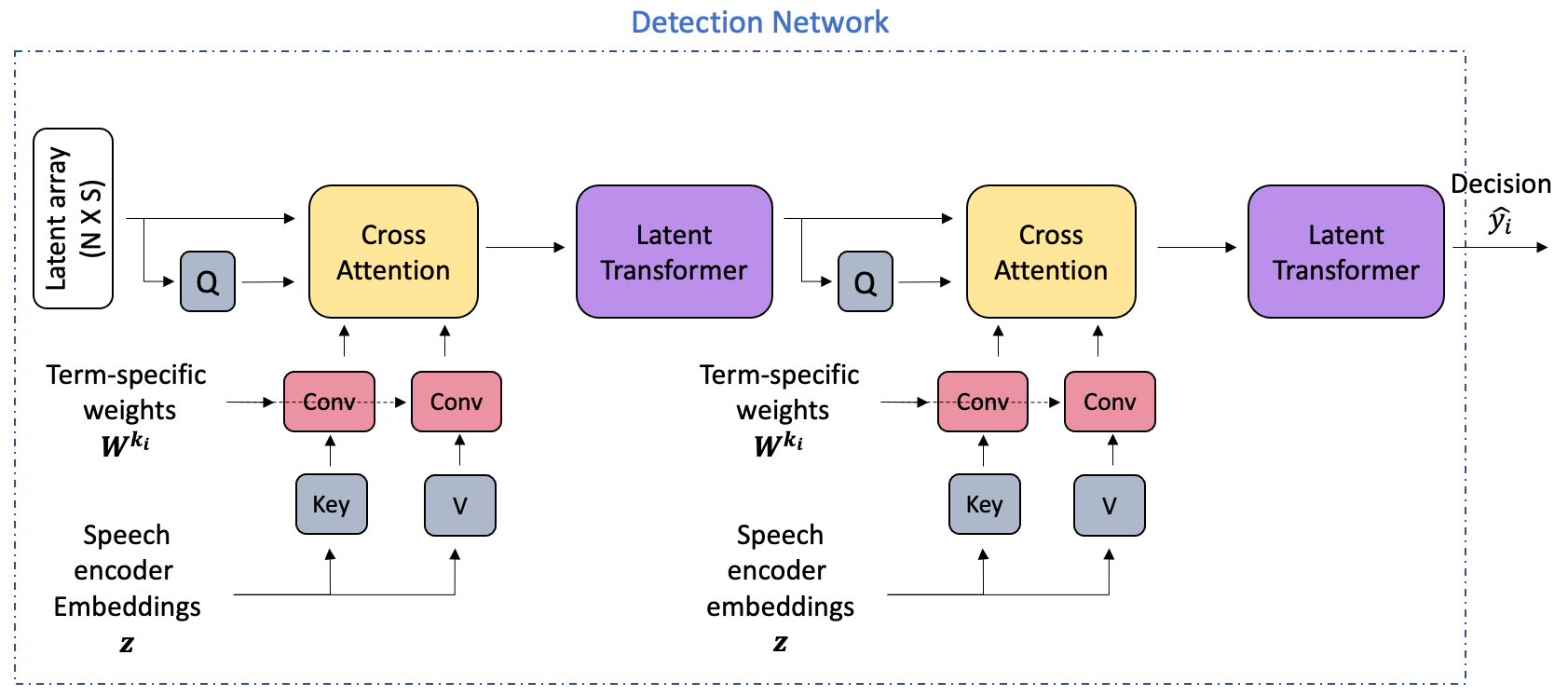}
        \caption{}
        \label{fig:sub2}
    \end{subfigure}
    \caption{\conv~model architecture: (a) \textbf{Overview}: The model consists of a \emph{Speech encoder} that generates a representation of the audio input. A separate \emph{Target keyword encoder} processes a keyword—defined by its sequence of characters—and produces corresponding weights for the \emph{Detection network}. Notably, the keyword encoder is trained only once and can generate weights for unseen keywords during inference. (b) \textbf{Detailed view} of the architecture of the Detection Network.}
\label{fig:architecture_all}
\end{figure*}

\section{Model}
\label{sec:model}

In the open-vocabulary KWS task, the model is provided with a speech utterance and must determine whether it contains a given target keyword, which is provided as a sequence of characters. It is important to note that while the training data includes a wide range of keywords, the open-vocabulary KWS model is designed to generalize and detect keywords it has not encountered during training. Our proposed model, called \emph{\conv}, consists of three sub-models: a speech encoder, a target keyword encoder, and a detection network. This architecture is illustrated in Figure \ref{fig:architecture_all}.

\subsection{Speech Encoder}

The speech signal,  denoted as $\x$, is characterized by a series of $T$ acoustic feature vectors, each of dimension $D$, sampled every 10 milliseconds (msec). The speech encoder maps the speech sequence $\x$ into a sequence of representations $\mathbf{z}$. The resultant sequence has a length of $B$, and each element is $M$-dimensional vector.

We implemented the the speech encoder using two types of architectures. The first is the encoder of the Whisper transformer model \cite{radford2023robust}, which was trained on 680,000 hours of audio. To ensure our model is suitable for small-footprint devices, we used the pre-trained \emph{tiny Whisper} variant, the most compact among Whisper models. It consists of 4 Transformer layers, each with 6 attention heads, and has a feature size of $M=384$, sampled every 20 msec. This configuration yields an encoder module that comprises approximately 7.6M parameters. When tiny Whisper is used as an encoder, our keyword spotting model is denoted as \emph{\wconv}.

An encoder with a 7.6M parameter is still fairly large for small-footprint devices. Therefore, we also tried a second type of encoder, the \emph{tiny Conformer} \cite{nishu2023flexible}. The tiny Conformer has 6 encoder layers, 4 attention heads, a convolution kernel of size 3, and a feature size of $M=144$, sampled every 40 msec. This yields a model that is comprised of 3.7M parameters. The model was trained with the CTC loss function as described in Section \ref{sec:experimental_setup}.
When tiny Conformer is used as an encoder, our keyword spotting model is denoted as \emph{\cconv}.


\subsection{ Target Keyword Encoder}

The second module receives the target keyword as a character sequence, denoted by $\kc$, as input and produces a set of weights, $\Wkc$. These weights are then utilized to guide the \emph{detection network} to identify the specified target keyword accurately.

This approach to hyper-tuning the detection network draws inspiration from Bluche \emph{et al.} \cite{bluche2020predicting} and Navon \emph{et al.} \cite{navon2023open}. In our work, similar to Bluche \emph{et al.} \cite{bluche2020predicting}, the encoder outputs weights  $\Wkc$ for a convolution layer that represents the target keyword $\kc$. These weights can be seen as the parameters of a \emph{keyword-specific matched filter}, designed to maximize the signal-to-noise ratio for the target signal (in this case, the textual target keyword) \cite{therrien1992discrete}. 

However, unlike Bluche \emph{et al.} \cite{bluche2020predicting}, where the target keyword encoder generates the entire set of weights for a detection network consisting of a single CNN layer, our detection network utilizes these weights to perform a keyword-specific convolutional operation within a Perceiver model. This guides the cross-attention weights in the Perceiver module, which then proceeds through the full Perceiver attention mechanism, as described in the following section. This process directs the output of the audio encoder toward the target keyword, enabling the detection of whether the keyword was spoken in the recording. 

Additionally, our method differs from Navon \emph{et al.} \cite{navon2023open}, which utilizes a target keyword encoder to output weights for a few normalization layers in a full Transformer with 2 layers—an approach that is less suitable for small-footprint devices. Our target keyword encoder has a character embedding layer with an embedding size of 161 and 4 LSTM layers with a hidden size of 256. In addition, it has 2 linear layers projecting the LSTM hidden state to $\Wkc$, the convolution dimension, which leads to 2.7M parameters. Small-footprint devices use a pre-defined set of target keywords. In our approach, adding a new target keyword \( \kc \) does not require retraining the target keyword encoder to produce the corresponding weights \( \Wkc \). Since these weights can be generated offline, the target keyword encoder itself does not need to be deployed on the device.
 
\subsection{Detection Network}

Our last module is the detection network; its inputs are the speech encoder embeddings $\z$ and the keyword-specific convolution weights $\Wkc$. We aim to maintain a small kernel size $|\Wkc|$ to accommodate a small-footprint device while ensuring that the convolution effectively captures the target keyword. In this work, we used a kernel size 16 with a dilation of 1 and a stride of 1 for both the Whisper (20 msec resolution) and the Conformer encoders (40 msec resolution). We employ depth-wise convolution to minimize the convolution weights, where each input channel is convolved with its own set of filters. This approach helps keep the convolution weights small while maintaining the efficacy of the convolution operation. As a result, we end with $|\Wkc|=1024$ (kernel size of 16 with 64 channels).

We also aim for the detection network to be compact yet effective in detecting target keywords. Therefore, our detection network uses the \emph{Perceiver} architecture \cite{jaegle2021perceiver}. The Perceiver's uniqueness comes from its cross-attention mechanism that maps a high-dimensional input to a fixed-dimensional latent bottleneck. Then, the Perceiver applies Transformer-style self-attention blocks to the latent bottleneck. The original Perceiver architecture maps the speech representation in two stages. First, it projects the speech representation $\mathbf{z}$ from a higher embedding space of dimension $M$ to a lower-dimensional space of dimension $N$, where $N \ll M$. Then, it maps the variable-length $B$ elements to fixed-length vectors of size $S$, independent of $B$.

In this work, the mapping occurs in three stages. First, the speech representation $\mathbf{z}$ is projected from a higher embedding space of dimension $M$ (384 for Whisper and 144 for Conformer) to a lower-dimensional space of dimension $N = 64$. Next, we incorporate our method, where the Perceiver implementation applies keyword-specific convolution (the \emph{Conv} element in Figure \ref{fig:sub2}). Finally, it maps the variable-length $B$ elements to fixed-length vectors of size $S = 16$, independent of $B$. We then evaluated the performance with 1 to 5 Perceiver layers, consisting of a cross-attention layer followed by a latent transformer.

\subsection{Loss Functions}
Both \wconv and \cconv models, each composed of a speech encoder, a target keyword encoder, and a detection network, were trained using the Binary Cross-Entropy (BCE) loss, defined as:
\begin{equation}  
    \mathcal{L}_{\text{BCE}} = -\frac{1}{E} \sum_{i=1}^{E}  y_i \log \hat{y}_i + (1 - y_i) \log (1 - \hat{y}_i)  ~,
\end{equation}  
where $E$ denotes the batch size,  $y_i$  is the binary label indicating the presence (1)  or absence (0) of the keyword, and $\hat{y_i}$ represents the predicted probability by the detection network of the keyword $k$ to being present in the audio sample $x$.

During training, the BCE loss is computed over the outputs of the detection network, and its gradients are backpropagated through the target keyword encoder and the detection network.

\section{Experiments}
\label{sec:experiments}

We begin by detailing the datasets used for training and evaluation, followed by a description of the experimental setup. The subsequent section presents the results.

\subsection{Dataset}
\label{sec:dataset}

Multiple datasets were used for the training and evaluation of our models. The first dataset, VoxPopuli \cite{wang-etal-2021-voxpopuli}, is a substantial multilingual speech corpus comprising approximately 1800 hours of transcribed utterances across 16 languages. This dataset was used for both training and evaluation. 
To prepare the VoxPopuli dataset for open-vocabulary KWS training, we adopted the approach outlined by \cite{navon2023open}, which involves generating negative examples in four different ways within each batch. The first method randomly selects a negative keyword from other target keywords in the batch. The second method creates a negative example by concatenating two different target keywords. The third method involves swapping one character in the target keyword with another. The fourth method selects the nearest negative keyword within the batch. During the evaluation phase, we employed similar negative sampling techniques, excluding the nearest negative keyword.

The second dataset, LibriPhrase \cite{shin2022learning}, is a recent benchmark for KWS based on the LibriSpeech \cite{panayotov2015librispeech} dataset. The authors provided the test part of the corpus. However, their released code for creating the training and validation parts lacks some parameters. Hence, we were able to generate approximately 200K examples (out of the 800K). LibriPhrase consists of two subsets: LibriPhrase Hard (LH) and LibriPhrase Easy (LE), each containing phrases varying in length from one to four words. This dataset was used for both the training and evaluation phases.

The LibriSpeech dataset \cite{panayotov2015librispeech} was primarily used for training the Conformer speech encoder. Due to the construction of the LibriPhrase test dataset from the train-others-500 subset, we limited our use of LibriSpeech to the train-clean-100 and train-clean-360 subsets for training purposes.

Lastly, we used three datasets for out-of-domain evaluation purposes.  The first dataset is Speech Commands V1 \cite{warden2018speech}, for which we followed the setup proposed by \cite{shin2022learning}, containing 10 different keywords. The second dataset is the test subset of the FLEURS dataset \cite{conneau2023fleurs}, which evaluates models on low-resource languages. The third dataset, the Wildcat Diapix corpus \cite{van2010wildcat}, consists of English dialogues produced by two speakers during an interactive task. Each speaker is looking at one of two versions of a picture showing a street scene. Using speech alone, the speakers must work together to find the differences between the two images. For keywords, we used a set of 20 words related to the 10 differences in the pictures. Critically, this dataset enables us to evaluate performance for both native English speakers (L1) and non-native English speakers (L2). We used all interlocutor types except for two speaker pairs that lacked corrected speech-to-transcript alignment. To prepare the Wildcat audio files for the keyword spotting task, each interactive session was segmented according to transcript alignment. Segments shorter than 0.5 seconds were discarded, while those longer than 30 seconds were split into two. This process yielded 7,745 audio files, with 5,948 files belonging to L2 English speakers and 1,797 files to L1 English speakers.

Table \ref{tab:dataset} presents an overview of all datasets used in this study, including the number of keywords used for training and testing.


\begin{table}[ht]
\centering
\resizebox{1\columnwidth}{!}{%
\begin{tabular}{lccccr}
\hline\hline
Dataset & Train & Eval & Language & \# Keywords & \# Keywords \\
        &       &      &          & (Training) & (Testing) \\
\hline
VoxPopuli\cite{wang-etal-2021-voxpopuli} & \checkmark & \checkmark & Multiple & Open & 13745 \\
LibriSpeech 100+360\cite{panayotov2015librispeech} & \checkmark & x & English & - & - \\
LibriPhrase\cite{shin2022learning} & \checkmark & \checkmark & English & Open & 4003 \\
SpeechCommands V1\cite{warden2018speech} & x & \checkmark & English & - & 10 \\
FLEURS\cite{conneau2023fleurs} & x & \checkmark & Multiple & - & 2045 \\
Wildcat Diapix\cite{van2010wildcat} & x & \checkmark & English & - & 20 \\
\hline\hline
\end{tabular}
}
\caption{Overview of datasets used. “Open” indicates training keywords were not fixed.}
\label{tab:dataset}
\end{table}

\subsection{Experimental Setup.}
\label{sec:experimental_setup}

The models discussed in Section \ref{sec:model} feature two types of speech encoders. The audio input for both \wconv and \cconv models consists of an 80-channel log-magnitude Mel spectrogram generated using a 25-msec window and a 10-msec stride. The Conformer speech encoder was pre-trained for the Automatic Speech Recognition (ASR) task on VoxPopuli \cite{wang-etal-2021-voxpopuli} and on the train-clean-100 and train-clean-360 subsets of the LibriSpeech dataset \cite{panayotov2015librispeech}. The Conformer was trained using three augmentation techniques: frequency masking, time masking, and additive noise. The additive noise was sourced from the MUSAN dataset \cite{snyder2015musan}, with the signal-to-noise ratio (SNR) randomly sampled between 5 and 15 dB. 
The augmentations were conducted using the TorchAudio library\footnote{https://pytorch.org/audio/2.0.1/}. The training utilized the CTC loss for 250 epochs, with Adam optimizer \cite{kingma2014adam}, a learning rate of 0.0001, and a batch size of 96. 
For the open-vocabulary KWS task, both \wconv and \cconv models were trained using Binary Cross Entropy (BCE) loss for a maximum of 250 epochs, employing a validation loss stop condition with patience of 40, an Adam optimizer, a batch size of 96, and a learning rate of 0.0001. We report standard metrics used for KWS evaluation, namely, the F1 score, Area Under the Curve (AUC), Equal Error Rate (EER), and the False Reject Rate at a False Acceptance Rate of 5\%, denoted as FRR@FAR5\%.

We conducted comparisons between our models and the latest open-vocabulary KWS techniques, including CED \cite{nishu2023flexible}, EMKWS \cite{nishu2023matching}, CMCD \cite{shin2022learning} and \cite{bluche2020predicting}, which are based on text-audio embedding and a fusion network between them. We also compare our models against the state-of-the-art approach proposed by \cite{navon2023open}. As the original model and source code were not publicly available, we re-implemented the method and trained it within our experimental framework. The results of this reimplementation are reported across all datasets under the name  \emph{AdaKWS-tiny*}. For the LibriPhrase dataset, we additionally report the results from \cite{navon2023open}, listed as AdaKWS-tiny.
We report the number of parameters for each model, accounting only for those that need to be stored on the device. This means we exclude the target keyword parameters from our model and from other models that use a similar architecture.
 

\section{Results}
\label{sec:results}

In this section, we present detailed experiments to evaluate our model's detection performance. We begin by assessing the models in fully trained configurations, followed by an evaluation of their generalization to out-of-domain scenarios. Finally, we provide ablation results to analyze the impact of the Whisper encoder depth on performance. Throughout the paper, the best results are highlighted in bold.

The model size is a key aspect of our work; therefore, we report results for configurations using 1 to 5 Perceiver layers in order to examine the trade-off between model size and performance. The number of Perceiver layers is indicated in parentheses (for example, \wconv (5) refers to the \conv model with the Whisper encoder as the audio encoder and 5 Perceiver layers). 


\subsection{Main Results}
\begin{table}[h]
    \caption{AUC, F1, and EER and \# of parameters on device, for VoxPopuli dataset.}
    \label{tab:voxpopuli}
\centering
    \resizebox{1\columnwidth}{!}{%
    \begin{tabular}{lcccc}
    \hline\hline
        model & AUC $\uparrow$ & F1 $\uparrow$ & EER $\downarrow$ & \# Params \\ \hline
        AdaKWS- tiny* \cite{navon2023open} & 98.01 & 91.93 & 6.724 & 13.5M\\
        \hdashline
        \wconv (5)& 98.54 & 94.27 & 5.430 & 10M\\
        \wconv (4)&  98.61 & 94.39& 5.442& 9.5M\\
        \wconv (3)&   98.55 & 94.34& 5.430 & 9M \\
        \wconv (2)&   98.41 & 94.00 & 5.845 & 8.6M\\
        \wconv  (1) & 98.48  & 94.12 & 5.699 & 8.1M\\
      
        \cconv (5)  & 98.78 & 94.67 & 5.076 & 6M\\
        \cconv (4) & \textbf{98.90} & \textbf{94.88} & \textbf{4.918} & 5.5M\\
        \cconv (3) & 98.76  & 94.72 & 5.541 & 5M\\
        \cconv (2) & 98.80 & 94.56 & 5.211 & 4.6M\\
        \cconv (1) & 98.66 & 94.52 & 5.308 & 4.2M\\

        \hline\hline
    \end{tabular}%
    }
  
\end{table}

We begin by evaluating our models on our open-vocabulary KWS adaptation of the VoxPopuli dataset. Table \ref{tab:voxpopuli} reports the AUC, F1, and EER scores across multiple languages. In this setup, only the detection network and the target keyword encoder were trained, while the speech encoders in both the \cconv and \wconv models remained frozen.

The results indicate that \cconv(4) outperforms other models on the VoxPopuli dataset, with \cconv(5) and \cconv(4) achieving very close results. Notably, \cconv(3) presents slightly worse results than the other \cconv models; however, its performance remains strong. Interestingly, the \conv~models based on the Whisper encoder, despite having more parameters, perform slightly worse than the \conv~models based on the Conformer encoder.
Moreover, the \wconv(3) and \wconv(5) models exhibit very similar performance, despite a difference of 1 million parameters. Although we were unable to replicate the full AdaKWS-tiny result on VoxPopuli, our results are close to what the authors reported (F1 = 92.8). Nonetheless, all of our models achieved better performance than the AdaKWS-tiny* model.

\begin{table*}
    \caption{F1 results on VoxPopuli dataset by language}
    \label{tab:voxpopuli_lang}
\centering
    \resizebox{1\textwidth}{!}{%
    \begin{tabular}{lcccccccccccccccccc}
    \hline\hline
        model & CS & DE &EN & ES & ET & FI & FR & HR & HU & IT & LT & NL & PL & RO & SK & SL & ALL\\ \hline
        \wconv (5)& 94.20 & 96.10 & 96.66 & 96.13 & 82.61& 90.50 & 96.28 & 90.30 & 90.92& 92.21 & 77.27& 91.44 & 95.92 & 94.09 & 93.33& 87.93 & 94.27\\
        \wconv (4)& 94.25 & 96.28 & 97.15 & 95.94 & \textbf{86.96} & 92.14 & 96.34 & 91.06 & 90.72 & 91.36 & 71.73 & 91.70 & 95.55 & 94.78 & 93.90 & \textbf{88.54}& 94.39\\
        \wconv (3)& 94.54 & 95.62& 97.15 &  96.14& 79.07 & 91.60& 96.24 & 88.95 &  91.76& 92.17 & 75.56 &  91.94& 95.30 & 94.31 & 93.58& \textbf{89.03}& 94.34\\
        \wconv (2)& 94.61 & 95.59&  96.88&  95.49& 69.57 & 89.35& 95.44 & 89.07 & 91.91 & 91.82 &  83.33& 92.00 &  94.84&  93.89& 94.23& 87.78& 94.00\\
        \wconv (1)& 94.25 & 94.92& 96.99 &  96.06& 82.61 & 89.71&  96.01  & 90.45 &  90.16&  \textbf{92.45}& 83.33 & \textbf{92.78} &  94.50& 95.00& 94.77& 85.00& 94.12\\
        \cconv  (5)&94.68 & 96.35 & 97.03 & 96.01 & 79.07 & \textbf{93.25} & 96.53 & 91.27 & 92.05& 90.97 & 85.11&  91.84& 95.92 &\textbf{95.35}& 95.53 &87.31 & 94.67\\
        \cconv  (4)& \textbf{95.62} & \textbf{96.42}& 97.42 &95.73  &  80.85&92.05 &96.40  &90.96  & \textbf{93.72} & 91.15 & 80.00 & 92.45 & \textbf{96.29} & 94.80 & \textbf{96.33}& 86.83& \textbf{94.88}\\
        \cconv (3)& 94.76 & 96.32& 97.65 & 96.18 & 80.85 & 93.13& 95.96 & \textbf{91.57} &92.83  &91.22  & 83.33 & 92.22 & 95.58 & 95.08 & 94.44&87.31 & 94.72\\
        \cconv (2)&  93.49& 95.82&  97.14& \textbf{96.30} & 81.82 &91.82 & 96.23& 91.37 &92.17  & 91.76 & \textbf{86.36} & 92.36 & 95.82 & 94.81 &94.64 &88.29 & 94.56\\
         \cconv (1)&  93.75& 96.16& \textbf{97.83} & 95.52& 79.07&91.25 & \textbf{96.58}& 90.76 & 91.48&  91.60&  81.82&  92.39&  95.83&95.13 &94.23 &86.06 & 94.52\\
        \hline\hline
    \end{tabular}%
    }
    
\end{table*}

As Table \ref{tab:voxpopuli} presents results across multiple languages, Table \ref{tab:voxpopuli_lang} reports the F1 scores for the VoxPopuli dataset, broken down by language.
The table reveals consistent performance across all languages, with no single model significantly outperforming the others. However, an intriguing observation is noted in the Lithuanian (LT) and Estonian (ET) languages, where the F1 scores are relatively low (71.73, 69.57). This discrepancy may be attributed to the relatively low amount of training examples for Lithuanian ($\sim$ 500) and Estonian ($\sim$ 800) compared to the other languages ($>$10k).

\subsection{Fine-tuning on New Data} 

\begin{table}
    \caption{AUC and EER comparison results on the LibriPhrase dataset's hard (LH) and easy (LE) test splits.}
    \label{tab:libri_phrase}
    \centering
    \resizebox{1\columnwidth}{!}{%
    \begin{tabular}{lcccccc}
    \hline\hline
     & \multicolumn{2}{c}{AUC $\uparrow$} & \multicolumn{2}{c}{EER $\downarrow$} &  \# Params\\[0.05cm]
         Model &  LE & LH & LE & LH \\
         \hline
        CMCD \cite{shin2022learning} & 96.7 & 73.58 & 8.42 & 32.9& UKN\\
        EMKWS \cite{nishu2023matching} & 97.83 & 84.21 & 7.36 & 23.36& 3.7M\\
        CED \cite{nishu2023flexible} & 99.84 & 92.7 & 1.7 & 14.4& 3.7M\\
        AdaKWS-tiny \cite{navon2023open} & 99.8 & 93.75 & 1.61 & 13.47 & 13.5M\\
        AdaKWS-tiny* \cite{navon2023open}  & 99.63 & 94.46 & 2.25  & 12.107& 13.5M\\
        \hdashline
         \wconv (5)& 99.90 & 95.6 & 1.22 & 10.97 & 10M\\
        \wconv  (4)& 99.88 & 95.57 & 1.28 &  11.01&9.5M\\
        \wconv  (3)& 99.86 &  95.39& 1.338 & 11.12 & 9M\\
        \wconv  (2)& 99.88 & 95.40 & 1.212 &  11.10 & 8.6M\\
        \wconv  (1)& 99.90 & 95.30 & 1.122& 11.53 & 8.1M\\
         
         \cconv (5)& 99.86 & 95.84 & 1.27 & 10.77& 6M\\
        \cconv (4)& 99.89 &  \textbf{96.07}&\textbf{1.08}  & \textbf{10.45} &5.5M\\
        \cconv  (3)& \textbf{99.91} & 95.84 & 1.10 & 10.72  &5M\\
        \cconv  (2)& 99.87 & 95.62 & 1.21 & 11.22&4.6M\\
        \cconv  (1)&99.90 & 95.96& 1.12& 10.82 & 4.2M\\
        \hline\hline
    \end{tabular}%
    }

\end{table}

It appears that all our models performed well on the test set of the VoxPopuli dataset. We now evaluate our models on the LibriPhrase dataset and compare their performance to additional models. We used models pre-trained on VoxPopuli and fine-tuned them for an additional epoch using our version of the LibriPhrase training set. For the \cconv models, all components of the \conv~architecture—including the speech encoder—are fine-tuned. In contrast, for the \wconv models, only the detection network and the target keyword encoder are fine-tuned, while the speech encoder remains frozen. This decision is based on the assumption that the Whisper encoder, having been trained on a sufficiently large and diverse dataset, does not require additional training. Table \ref{tab:libri_phrase} presents the AUC and EER results.

It is clear that the \cconv(3) and \cconv(4) models outperform all other models. Moreover, both the \cconv and \wconv models surpass all baseline models across all metrics, highlighting the high effectiveness of both our Conformer-based and Whisper-based speech encoders for this task. Notably, the \cconv(1) model, with a parameter count similar to that of EMKWS \cite{nishu2023matching} and CED \cite{nishu2023flexible}, still outperforms these models across all metrics.

Interestingly, AdaKWS-tiny outperforms AdaKWS-tiny* on the LE portion of LibriPhrase, but this trend reverses on the LH portion. This discrepancy might be due to differences in the training setup, as we don't have access to the exact training code and model weights of AdaKWS-tiny.

Furthermore, the \wconv models achieve superior results compared to all models except the \cconv models, even when the encoder is frozen, underscoring the exceptional performance of the whisper encoder.

\subsection{Out-Of-Domain Results} 
We further evaluate our models on several out-of-domain datasets, none of which were used to train the \wconv and \cconv models.
\subsubsection{Fleurs} 

\begin{table}[h!]

    \caption{F1 results for out-of-domain performance for low-resource languages on Fleurs dataset.}
    \label{tab:fleurs}
    \centering
    \resizebox{1\columnwidth}{!}{%
    \begin{tabular}{lccccc}
    \hline\hline
        model & Icelandic & Maltese & Swahili & Uzbek & Overall \\ \hline

        AdaKWS- tiny* \cite{navon2023open} & 68.18&  \textbf{67.45}& 75.49 &  60.62 & 66.85\\
        \wconv (5)& 61.54&  64.55&  \textbf{78.57}&  60.49& 66.40\\
        \wconv (4)& 59.46&  53.64&  74.68 &  54.36& 59.27\\
        \wconv (3)& 47.06 &55.30&69.07 & 45.80 & 54.96\\
        \wconv (2)& 51.43 & 58.29 &74.16 & 51.98& 59.91\\
        \wconv (1)&  50.00 & 63.82& 72.62& 54.98 & 62.84\\
         \cconv (5)& 64.86& 54.34 & 68.89 &  58.68 & 59.60\\
        \cconv (4)& 68.29 & 65.30 & 69.32 & \textbf{67.45}  &  \textbf{67.16}\\
        \cconv (3)& 57.14 & 58.10 & 72.92 &  62.86&  63.27 \\
        \cconv (2)&\textbf{ 70.00} &  62.27& 70.00 & 62.47 & 64.61  \\
        \cconv (1)& 60.47 & 63.41  &  69.57 & 59.83 & 63.64\\

        \hline\hline
    \end{tabular}%
    }
   
\end{table}

We begin by evaluating our models on low-resource languages, considering them out-of-domain in two aspects: the dataset was not used for training, and these languages were absent from the training data. This evaluation is conducted using the Fleurs dataset, with F1 scores reported in Table \ref{tab:fleurs}.

It is notable that AdaKWS* was one of the top-performing models in this case, which contrasts with its performance on previous datasets.
Interestingly, all the \wconv models except \wconv(5) show large degradation in F1 scores. This is surprising, as we would expect the whisper-base models to perform better. Additionally, \wconv(1), which has the fewest parameters among the \wconv models, performs quite well.
Contrary to this, the \cconv models outperform the \wconv models, with \cconv(4) again achieving the best results. Similar to the \wconv models, larger Perceiver depth does not necessarily lead to better performance.

\subsubsection{Speech Command} 

\begin{table}
    \caption{AUC, EER, and  FRR@ FAR5\% and \# of parameters on device, for Speech Command dataset.}
    \label{tab:speech_command}
\centering
    \resizebox{1\columnwidth}{!}{%
    \begin{tabular}{ccccc}
    \hline \hline
        model & AUC $\uparrow$ & EER $\downarrow$  & FRR@ FAR5\% $\downarrow$ & \# Params\\
        \hline
         Detection filters \cite{bluche2020predicting}& - & - & $\sim$ 36 & 210K\\
         Detection filters fine-tuned \cite{bluche2020predicting}& - & - & $\sim$ 25 & 210K\\
        CED \cite{nishu2023flexible} & 93.94 & 13.45 & - & 3.7M\\
        \hdashline
        \wconv (5) & \textbf{99.05}& \textbf{5.056 }& \textbf{5.064 }& 10M\\
        \wconv (4) &  98.78&  5.670& 5.960& 9.5M\\
        \wconv (3) & 98.66 & 6.154 & 6.739 & 9M\\
        \wconv (2) & 98.61 & 6.193 & 7.168& 8.6M\\
        \wconv (1) & 98.05& 7.205 & 8.220& 8.1M\\

     \cconv (5)  & 97.75 & 8.554 &  10.91 & 6M\\
        \cconv (4)  & 96.89 & 8.877 & 12.54& 5.5M\\
        \cconv (3)  & 96.48 & 9.896 & 13.60 & 5M\\
        \cconv (2)  & 96.90 & 9.351 & 14.06& 4.6M\\
       \cconv (1)  & 97.25 & 9.716& 13.13& 4.2M\\

         \hline \hline
    \end{tabular}%
    }
    
\end{table}

Next, we perform out-of-domain evaluation using the Speech Commands dataset, which differs significantly from VoxPopuli in both acoustic environment and lexical content and is widely used for keyword-spotting tasks. We selected 10 keywords from the dataset, as done in \cite{nishu2023flexible}. Since the audio files in the Speech Commands dataset are very short, about 1 second in length, we evaluated our model using the version fine-tuned on the LibriPhrase dataset, which also consists of short audio clips. The results are presented in Table \ref{tab:speech_command}.

All of our models outperform the Detection Filter model \cite{bluche2020predicting}, even when the latter is fine-tuned on the Speech Command dataset. This is expected, given that our models have an order of magnitude more parameters. However, even when compared to the CED model \cite{nishu2023flexible}, our \cconv and \wconv models still perform better. On the Speech Command dataset, the trend shifts, with our \wconv models outperforming the \cconv models. Specifically, \wconv(5) emerges as the best model overall, while \cconv(5) is the top performer among the \cconv models.

\subsubsection{Wildcat Diapix Corpus} 
\begin{table}[h]  
     \caption{AUC, EER, and  FRR@ FAR5\% for Wildcat dataset by speaker type.}
      \label{tab:wildcat}
     \centering
    \resizebox{1\columnwidth}{!}{%
    \begin{tabular}{cccccc}
    \hline\hline
       subset & model & AUC $\uparrow$  & EER $\downarrow$ & FRR@ FAR5\% $\downarrow$ \\ \hline    
       \multirow{10}{*}{L1 English} &\wconv (5) &  \textbf{98.11}& \textbf{6.962 } & \textbf{8.933} \\
        &\wconv (4) & 97.55 & 7.540  & 9.397\\
        &\wconv (3) &  97.25& 8.932 & 11.949 \\
        &\wconv (2) & 97.57 & 7.492 & 9.397\\
        &\wconv (1) & 95.67  & 10.56  & 14.27\\
        &\cconv (5) & 94.17  &  12.30  & 15.55\\
        &\cconv (4) & 94.27 & 13.01  & 18.33 \\
        &\cconv (3) &  95.86 &  11.02 & 15.08 \\
        &\cconv (2) &  96.54 &  9.821 & 14.15\\
        &\cconv (1) & 93.91&  13.34  & 17.98\\
\hdashline
        \multirow{10}{*}{L2 English} &\wconv (5) &  \textbf{96.58} & \textbf{9.545} & 14.11  \\
        &\wconv (4) & 95.28 & 10.93 & 15.64 \\
        &\wconv (3) & 95.95& 9.996  & 12.95\\
        &\wconv (2) & 95.98 &9.633  & \textbf{12.63} \\
        &\wconv (1) & 95.58 & 10.53  & 13.08\\
        &\cconv (5) &  93.11  &  13.44 & 20.21\\
        &\cconv (4) & 93.56  &  13.92 & 19.94\\
        &\cconv (3) & 95.05 &  12.35 & 19.09 \\
        &\cconv (2) & 95.11 &  12.26 & 18.19  \\
        &\cconv (1) & 93.76 &  14.15 & 20.43\\
    \hline\hline
    \end{tabular}   %
    }
    
\end{table}

We conclude our out-of-domain evaluation by testing the model on the Wildcat Diapix corpus, which, in contrast to the VoxPopuli dataset, features spontaneous speech and includes many speakers for whom English is a second language (L2).\footnote{This is the first time this dataset has been used for the keyword spotting task, and as a result, there are no existing results from other methods on this dataset. We will share our evaluation script to allow this dataset to be used to compare L1 vs. L2 results in future research.}.

Results are shown in Table \ref{tab:wildcat} and show that the system performs at a high level for both L1 and L2 speakers. As observed with the Speech Command dataset, the \wconv models outperform the \cconv models on the Wildcat dataset. Specifically, \wconv(5) achieves the best results for L1 English speakers and shows top performance in AUC and EER measures for L2 speakers. In contrast, the \cconv models generally produce lower results across the dataset. Although there is a slight drop in performance on the L2 speaker subset for both \wconv and \cconv models, the overall accuracy remains high,despite the fact that the training data did not include English speech from L2 speakers. These results demonstrate the system’s strong generalization capabilities in this out-of-domain setting.


\subsection{Ablation Study}\label{sec:ablation}

\begin{table}[h]
    
    \caption{Impact of Model Depth on Performance: The table presents AUC, F1, and EER scores, along with the number of parameters on the device, for models with varying numbers of Perceiver and Whisper Encoder layers. Here, \#E Layers denotes the number of Transformer encoder layers, while \#D Layers represents the number of Perceiver layers.}
    \label{tab:depth}
    \centering
    \resizebox{1\columnwidth}{!}{%
    \begin{tabular}{ccccccc}
    \hline\hline
         model &  \# E Layers&  \# D Layers &  AUC $\uparrow$ & F1 $\uparrow$ & EER $\downarrow$ & \# Params \\ \hline
        \wconv  & 4 & 5 & 98.54 & 94.27 & \textbf{5.430} & 10M\\
        \wconv & 3 & 5 & 98.23 & 93.6& 6.113 & 8.2M\\
        \wconv & 2 & 5 & 97.59 & 92.51& 7.431 & 6.4M\\
        \hdashline
        \wconv & 4 & 4 & \textbf{98.61} & \textbf{94.39} & 5.440 & 9.5M\\
        \wconv & 3 & 4 & 98.13 & 93.26 & 6.504 & 7.7M\\
        \wconv  & 2 & 4 & 97.67 & 92.46& 7.236 & 6M\\
        \hdashline
        \wconv & 4 & 3 & 98.55 & 94.34& \textbf{5.430} & 9M\\
        \wconv & 3 & 3 & 98.09 & 93.67 & 6.480 & 7.3M\\
        \wconv & 2 & 3 & 97.61 & 92.78 & 7.236 & 5.5M\\
          \hdashline
        \wconv & 4 &  2 & 98.41 & 94.00 & 5.845 & 8.6M\\
         \wconv & 3 &  2 & 98.24 & 93.80 & 6.040 &6.8M \\
          \wconv & 2 &  2 & 97.58 & 92.68 & 7.322 & 5M\\
           \hdashline
        \wconv & 4 &  1 & 98.11 & 93.44 & 6.345 & 8.1M\\
         \wconv & 3 &  1 & 97.89 & 93.09 & 6.663 &6.3M\\
          \wconv & 2 &  1 & 97.75 & 92.95 & 7.016 & 4.6M \\
        \hdashline
        \hline\hline
    \end{tabular}%
    }
    
\end{table}

Given the importance of model size in our work, we investigate the effect of using different numbers of Transformer layers from the tiny Whisper Transformer Encoder to assess their significance on the \wconv model performance. This ablation study was conducted on our open-vocabulary KWS adaptation of the VoxPopuli dataset, with results presented in Table \ref{tab:depth}. 

The results show that the best performance across various Perceiver layer configurations is achieved when the model utilizes all four transformer layers of the tiny Whisper Transformer Encoder. Models with these four transformer layers consistently demonstrate strong performance. However, if the user is willing to trade some accuracy for a smaller model, it is still possible to achieve very good performance without using all the encoder layers.

\section{Analysis}
We conclude the paper with a noise robustness study and an analysis of the Perceiver's cross-attention mechanism. 

\subsection{Noise Analysis}

\begin{figure*}[h]
    \centering
    \begin{subfigure}[b]{0.49\textwidth}
        \centering
        \includegraphics[width=\textwidth]{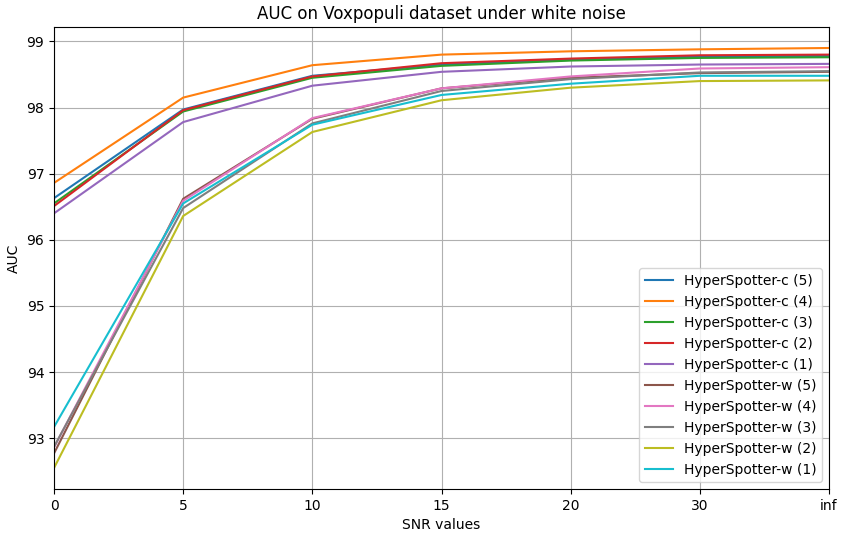}
        \caption{}
        \label{fig:white_auc}
    \end{subfigure}
    \hfill
    \begin{subfigure}[b]{0.49\textwidth}
        \centering
        \includegraphics[width=\textwidth]{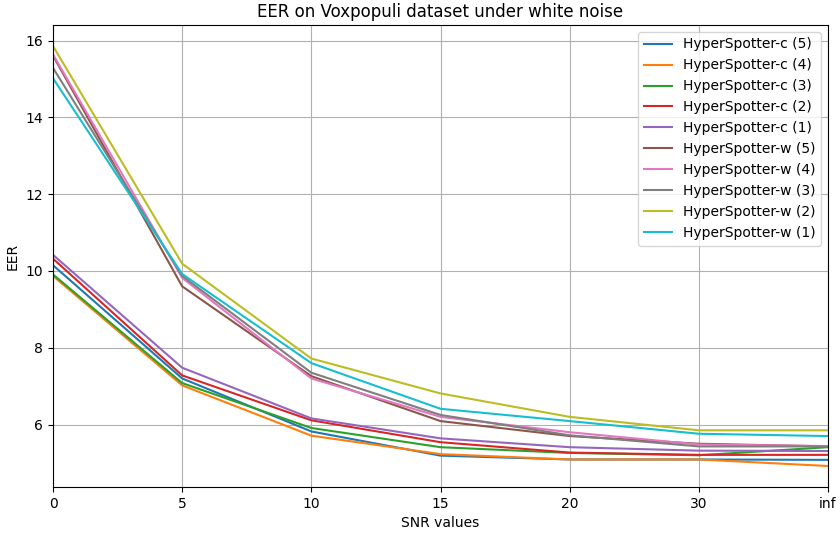}
        \caption{}
        \label{fig:white_eer}
    \end{subfigure}
    \begin{subfigure}[b]{0.49\textwidth}
        \centering
        \includegraphics[width=\textwidth]{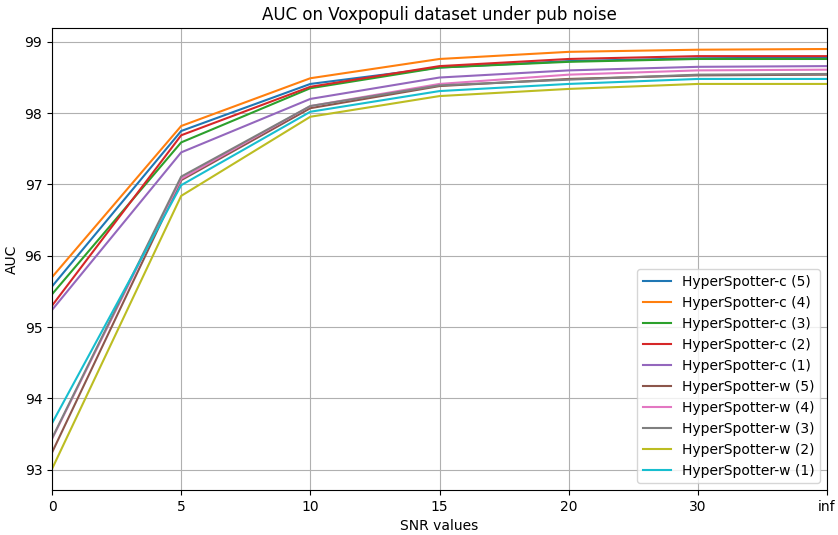}
        \caption{}
        \label{fig:pub_auc}
    \end{subfigure}
    \hfill
    \begin{subfigure}[b]{0.49\textwidth}
        \centering
        \includegraphics[width=\textwidth]{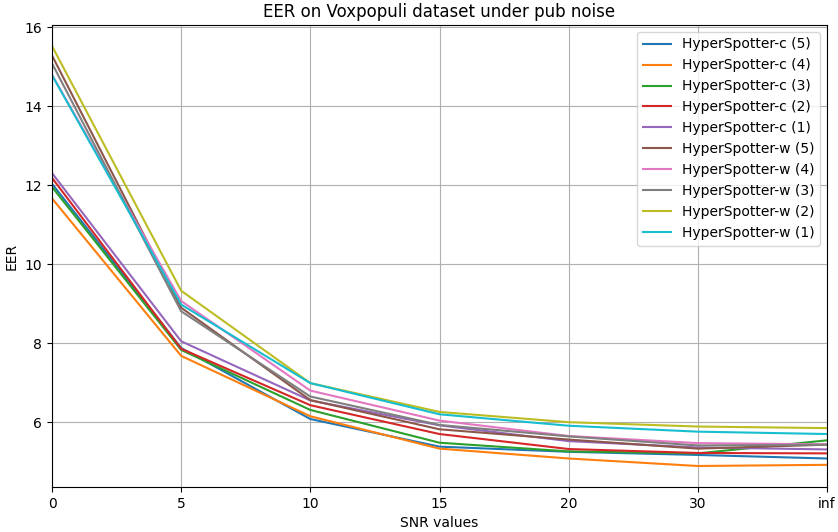}
        \caption{}
        \label{fig:pub_eer}
    \end{subfigure}
    \caption{AUC and EER measurements on the Voxpopuli dataset across different SNR levels under pub and white noise conditions. (a) AUC under white noise, (b) EER under white noise, (c) AUC under pub noise, (d) EER under pub noise.}
    \label{fig:white_pub_noise}
\end{figure*}

Our models have shown strong performance across various datasets, but we also aim to assess their effectiveness in noisy environments. Figure \ref{fig:white_pub_noise} presents the EER and AUC scores of our model on the Voxpopuli dataset under white noise and pub noise conditions, as described in the Whisper paper \cite{radford2023robust}, across different SNR levels.

Consistent with the other datasets, \cconv(4) delivers the best performance in both AUC and EER across all SNR levels under pub and white noise conditions. Additionally, \wconv(5) also achieves very strong results.

Overall, all models perform well across different SNR levels. Surprisingly, the \cconv models outperform the \wconv models. Although the \cconv Conformer encoder was pre-trained with additional noise, we initially expected the \wconv models to perform better, given that their encoder was trained on significantly more data. Nevertheless, the \wconv models still achieve very strong results.
As seen in previous results, \cconv(4) consistently achieves the best performance in both AUC and EER across all SNR levels under pub and white noise conditions. Additionally, \wconv(5) also performs very well.

\subsection{Cross Attention Analysis}

\begin{figure}[h]
    \centering
    \includegraphics[width=0.9\linewidth]{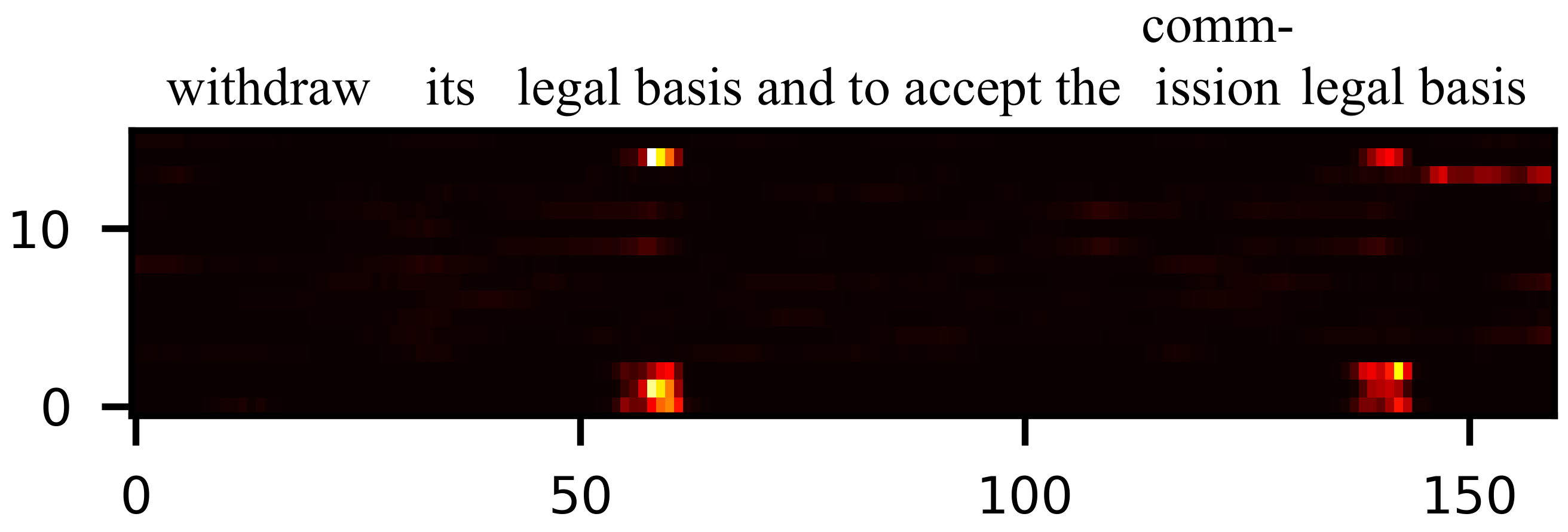}
    \caption{Heatmap of the \wconv(5) Perceiver's cross-attention for the target keyword ``legal basis'': showing the probability distribution of the Perceiver's fixed dimension $S$ (y-axis) across various time frames (x-axis). Higher probability scores correspond to the target keyword's locations. Darker is lower, and brighter is higher.}
    \label{fig:heatmap}
\end{figure}
We aim to explore our system to confirm whether our keyword-specific convolution functions as a matched filter, meaning that the Perceiver's cross-attention mechanism effectively highlights the target keyword's location in the audio. Figure \ref{fig:heatmap} depicts a heatmap of the \wconv(5) Perceiver's cross-attention for the target keyword ``legal basis''. In this heatmap, the y-axis represents the Perceiver's fixed dimension $S$, while the x-axis corresponds to the speech encoder embedding time frames (20 msec resolution). Notably, higher probability scores align with the two instances where the target keyword is spoken.

\subsection{In-Vocabulary and Out-of-Vocabulary Keywords}

One important aspect of the open-vocabulary KWS task is the question of performance related to In-Vocabulary (IV) and Out-of-Vocabulary (OOV) keywords. Generally, open-vocabulary KWS systems perform much better for IV setups than for OOV setups. In our work, evaluating the model's performance for these setups is not possible. Since during training we dynamically generate the negative keywords and randomly select positive keywords, we cannot determine which keywords were included in the training and which were not. However, since the keyword generation process during evaluation mirrors that of training, we assume that some of the evaluation keywords were not seen during training. Therefore, the distinction between in-vocabulary (IV) and out-of-vocabulary (OOV) keywords remains ambiguous. It is important to note that the selection of negative and positive keywords during evaluation was deterministic.

\section{Conclusion}
\label{sec:conclusion}

In this study, we introduce \conv, a novel end-to-end system designed for spoken keyword detection on small-footprint devices. The proposed model combines various speech encoders with a hyper-network that generates keyword-specific convolutional weights, effectively acting as a matched filter tailored to each target keyword. We evaluate two variants of the model, \wconv and \cconv, both of which achieve state-of-the-art detection accuracy across several benchmark datasets, including VoxPopuli, LibriPhrase, and Speech Commands. In addition, our experiments on the Wildcat dataset—featuring non-native (L2) English speakers—demonstrate the strong generalization capabilities of our approach in out-of-domain conditions.

Our results show a consistent performance distinction between the two model variants: the \cconv models tend to perform better on datasets used for training and fine-tuning, while the \wconv models exhibit superior robustness in out-of-domain scenarios. We also analyze the effect of model depth and find that the optimal number of Perceiver layers depends on the choice of encoder. Specifically, four layers are generally ideal when using a Conformer-based encoder, while five layers yield the best results with the Whisper encoder.

Looking ahead, we plan to explore additional types of speech encoders and detection architectures to further reduce the model’s size and computational footprint, while maintaining or improving performance. 

\section{Acknowledgements}

This work is supported by the National Institutes of Health under grant MH134369. Y. Segal-Feldman is sponsored by the Ministry of Science \& Technology, Israel.

\bibliographystyle{IEEEtran}
\bibliography{refs}

\end{document}